\documentclass[journal=nalefd,manuscript=letter]{achemso}

\usepackage{upgreek}
\usepackage[version=3]{mhchem} 
\usepackage[T1]{fontenc}       
\usepackage{color}
\usepackage{multirow}
\usepackage{makecell}
\usepackage{amssymb}
\usepackage{amsmath}
\usepackage[normalem]{ulem}
\graphicspath{{figures/}}

\author{Cedric Klinkert$^{\dagger}$}
\author{\'Aron Szab\'o$^{\dagger}$}
\author{Christian Stieger$^{\dagger}$}
\author{Davide Campi$^{\ddagger}$}
\author{Nicola Marzari$^{\ddagger}$}
\author{Mathieu Luisier$^{\dagger}$}
\email{mluisier@iis.ee.ethz.ch}
\affiliation{$^{\dagger}$ Integrated System Laboratory, ETH Zurich,
  CH-8092 Zurich, Switzerland\\ $^{\ddagger}$ Theory and Simulation of
  Materials (THEOS) and National Centre for Computational Design and
  Discovery of Novel Materials (MARVEL), \'Ecole Polytechnique
  F\'ed\'erale de Lausanne, CH-1015 Lausanne, Switzerland}

\title{2-D materials for ultra-scaled field-effect transistors:
hundred candidates under the \textit{ab initio} microscope}

\begin{document}

\begin{abstract}
Thanks to their unique properties single-layer 2-D materials appear as  
excellent candidates to extend Moore's scaling law beyond the
currently manufactured silicon FinFETs. However, the known 2-D
semiconducting components, essentially transition metal 
dichalcogenides, are still far from delivering the expected
performance. Based on a recent theoretical study that predicts the
existence of more than 1,800 exfoliable 2-D materials, we investigate
here the 100 most promising contenders for logic applications. Their
``current vs. voltage'' characteristics are simulated from
first-principles, combining density-functional theory and
advanced quantum transport calculations. Both $n$- and $p$-type  
configurations are considered, with gate lengths ranging from 
15 down to 5 nm. From this unprecedented collection of electronic
materials, we identify 13 compounds with electron and hole currents
potentially much higher than in future Si FinFETs. The resulting
database widely expands the design space of 2-D transistors and
provides original guidelines to the materials and device engineering 
community.

\textbf{Keywords:} 2-D materials, \textit{ab initio} device
simulation, next-generation field-effect transistors, performance
comparison, materials and device parameters
\end{abstract}

For more than five decades the functionality of all electronic
devices has not stopped growing driven by the continuous
miniaturization of their active components, the transistors. As a
consequence, the active region of the currently manufactured silicon
Fin field-effect transistors (FinFETs) does not exceed a couple of
nanometers along all directions. This renders their fabrication
extremely difficult and gives rise to strong quantum mechanical
effects such as inter-/intra-band tunneling that might negatively
impact their switching behavior. Furthermore, with the resulting
ultra-short gate lengths, $L_g\leq$20 nm, maintaining good
electrostatic properties has become really hard, even with
multi-gate architectures. 

Different options are therefore considered to extend Moore's scaling 
law (More-Moore) or go beyond it (More-than-Moore) \cite{kuhn}. On the
geometry side, nanowire \cite{nw} and nanosheet \cite{ns} FETs are
attracting a lot of attention because of their excellent
electrostatics and relative straightforward fabrication processes. On
the material side, logic switches based on III-V compound
semiconductors \cite{alamo}, germanium \cite{lugstein}, and carbon
nanotubes \cite{dai} have long been investigated to replace silicon,
all operating closer to their ballistic limit than Si.

The advent of graphene in 2004 \cite{novoselov} brought another
serious contender into this disputed race, two-dimensional (2-D)
materials. The absence of band gap did not allow graphene to emerge as
a rival to Si, but its discovery paved the way for other 2-D crystals,
starting with MoS$_2$ in 2011 \cite{kis}. Other members of the
transition metal dichalcogenide (TMDs) family have since then been
used as single- or few-layer transistor channels
\cite{wse2,ws2,mote2,mose2,hfse2,res2}. TMDs, like other 2-D
materials, benefit from dangling-bond-free surfaces and from an 
excellent electrostatic integrity provided by their atomic-scale
thickness. Sub-threshold swings close to the ideal value of 60
mV/dec at room temperature (RT) can be achieved with a single-gate
contact. Moreover, TMDs exhibit reasonable mobilities, up to 50-100
cm$^2$/Vs at RT, and band gaps compatible with logic operations 
($E_g$>1 eV). Taking advantage of these peculiar features, an
ultra-scaled MoS$_2$ transistor with a carbon nanotube of diameter
$d$=1 nm as gate electrode was demonstrated \cite{1nm}.  

While impressive, these results are still not sufficient to outperform
Si FinFETs. On one hand, there remain several technology issues to
address, notably the high contact resistance of monolayer TMDs
\cite{contact}. On the other hand, no drive current superior to 0.7
mA/$\mu$m has ever been reported \cite{pop}. This shortcoming 
can be partly attributed to the large effective masses of these
materials in comparison to Si, which limits their carrier injection
velocities and their ON-state current. Hence, 2-D components with
potentially better transport properties than TMDs have come under the
spotlight, e.g. black phosphorus (P$_4$) \cite{BP}, silicene
\cite{silicene}, germanene \cite{germanene}, antimonene
\cite{antimonene}, InSe \cite{inse}, Bi$_2$O$_2$Se \cite{bi2o2se}, and 
dozens of others \cite{fiori,franklin,r0,r1,r2}.

An intense research activity on novel 2-D materials is on-going at
the theoretical level too. A recent study based on density-functional
theory (DFT) predicted the existence of more than 1,800 exfoliable 2-D
candidates, among which several could play a major role as future,
post Si FinFET transistors \cite{mounet}. Since most of them have not
yet been experimentally isolated in their monolayer form, \textit{ab
  initio} device simulation is key to unlock their potential as
ultimate FETs. Up to now, the modeling effort has mainly focused on
known 2-D materials such as TMDs \cite{register,luisier,yoon},
P$_4$ \cite{klimeck}, Bi$_2$O$_2$Se \cite{quhe}, monochalcogenides
\cite{lake}, group IV \cite{wvdb} and V \cite{fiori2} monolayers, as
well as on more exotic ones, e.g. Tl$_2$O \cite{heine} whose electron
mobility was studied.

Here, we significantly expand the design space and report the largest
to date collection of simulated transistor characteristics together
with the associated material parameters for electronic applications. The
objective is to provide the community with a comprehensive atlas of
2-D materials capable of challenging Si FinFETs and to inspire the
work of engineers exploring next-generation 2-D FETs. With this
goal in mind, the 100 most promising entries from the database of
Ref.~\cite{mounet} are examined in their $n$- and $p$-type
configuration, with gate lengths ranging from 15 nm down to 5 nm,
assuming a simple, but realistic device geometry, beyond-ballistic
transport, perfectly ohmic contacts, and ideally insulating dielectric
layers. All simulations are performed at the DFT level with a
sophisticated \textit{ab initio} quantum transport solver that relies
on unique physical and numerical innovations \cite{omen}. We determine the
intrinsic transport properties of each 2-D material before comparing
them to each other. The detrimental influence of contact resistances
\cite{contact}, interactions with charged impurities \cite{fischetti},
structural defects \cite{defect}, and surface optical phonon
scattering \cite{jena} is ignored because all these effects are
expected to diminish as crystal quality and fabrication techniques
improve. 

In particular, we show that 13 2-D materials, among them 4 completely
new ones, stand out, delivering an ON-state current larger than 3
mA/$\mu$m at a fixed OFF-state magnitude of 0.1 $\mu$A/$\mu$m, both as
$n$- and $p$-type single-gate transistors. Three of these even have
ON-currents above 4 mA/$\mu$m in their $n$- and $p$-type FET
configuration. For each simulated 2-D compound, the electron/hole
``current vs. voltage'' characteristics, ``sub-threshold swing
vs. gate length'', unit cell, bandstructure, injection velocity,
transport and density-of-states effective masses, as well as a novel
physically relevant quantity called ``pass factor'' are presented in
the Supporting Information (SI). A detailed analysis of these data
follows.

\section{Results and Discussion}
\textbf{Device Test Structure.} As benchmark, the single-gate
transistor structure of Fig.~\ref{fig:1} is considered, as its
fabrication is fully compatible with standard CMOS processes. Its
design is loosely based on the specifications from the 2018 edition of
the International Roadmap for Devices and Systems (IRDS) \cite{irds}
for the year 2025, the first technology node (2.1 nm) at which 2-D
materials could eventually be inserted. The gate length is initially
set to $L_g$=15 nm (IRDS: 14 nm), the supply voltage to $V_{DD}$=0.7 V
(IRDS: 0.65 V), and the equivalent oxide thickness to $EOT$=0.6
nm. The source and drain extensions are doped with a donor/acceptor
concentration $N_{D/A}$=5e13 cm$^{-2}$, which is required to ensure an
excellent electrostatic control and access to the intrinsic properties
of 2-D FETs. Such values might be difficult to reach experimentally,
but they are definitively attainable \cite{pop,suh}. The 2-D materials
investigated are embedded between a SiO$_2$ box of thickness
$t_{box}$=20 nm and relative permittivity $\epsilon_R$=3.9 and a
HfO$_2$ dielectric layer with $t_{ox}$=3 nm and $\epsilon_R$=20 (see
Fig.~\ref{fig:1}). These oxide regions do not participate in the
electronic structure calculations, but they are nevertheless included
indirectly through the solution of Poisson's equation. 

Every simulation starts by constructing the primitive unit cell of the
selected 2-D material, which should have a band gap $E_g>$1 eV to
mitigate interband tunneling leakage. It continues with a relaxation
of the atomic positions and a computation of the corresponding
electronic structure; we use the Vienna Ab initio Software
Package (VASP) \cite{vasp}. Next, the plane-wave outputs of VASP are
transformed into maximally localized Wannier functions (MLWFs)
\cite{marzari} via the wannier90 code \cite{wannier90}. The assembled
MLWF Hamiltonian matrix is upscaled to match the targeted device
dimensions \cite{szabo1}. Finally, the \textit{I-V} characteristics of
the system are computed from first-principles with the OMEN quantum
transport solver \cite{omen}. The latter solves the Schr\"odinger and
Poisson equations self-consistently in the Non-equilibrium Green's
Function (NEGF) formalism, taking the produced tight-binding-like MLWF
Hamiltonian matrices as inputs. Electron-phonon interactions are
accounted for through a fictitious scattering model that eliminates
non-physical effects, e.g. negative differential resistance (NDR)
\cite{register}, and guarantees convergence of the electronic
currents up to the desired gate-to-source voltage $V_{gs}$=$V_{DD}$. 
More details about the modeling approach can be found in the Methods
Section. As it is the same for the 100 different 2-D channel materials
simulated in this work, a direct comparison of their device properties
is possible, opening up new avenues for 2-D electronics. 

\textbf{Performance Analysis.} The IRDS expects individual
high-performance (HP) Si FinFETs to reach ON-state currents
$I_{ON}$=1.67 mA/$\mu$m by 2025, when neglecting the influence of the
contact resistances, as assumed here. At the same time carrier
injection velocities $v_{inj}$=1.46e7 cm/s and sub-threshold swings
$SS$=80 mV/dec should be achieved. Figures \ref{fig:2}a and
\ref{fig:2}b report the \textit{I-V} characteristics of five of the
best 2-D FETs that we simulated at $L_g$=15 nm: P$_4$, GeS, 
As$_4$, Ag$_2$N$_6$, and SiH (silicane). They all deliver $I_{ON}>$3
mA/$\mu$m at $I_{OFF}$=0.1 $\mu$A/$\mu$m, in their $n$- and $p$-type
configurations, and could thus outperform the projected Si FinFET
currents by a factor $\sim$2 and monolayer MoS$_2$ by a factor
$>$3. In total we find 13 2-D materials that arrive at the same
performance level, as shown in Fig.~\ref{fig:2}c and summarized in
Table \ref{tab:1}. None of the conventional TMDs (MX$_2$ with M=Mo or
W and X=S, Se, or Te) belongs to this group, contrary to black
phosphorus (P$_4$) with transport along its $\Gamma$-$X$ axis, which
displays the highest ``$n$-type $I_{ON}$ vs. $p$-type $I_{ON}$''
combination. It should be noted that several 2-D candidates (39)
exhibit an ON-current larger than 3 mA/$\mu$m in their $n$-type form,
e.g. P$_8$Si$_4$ (5.43 mA/$\mu$m), As$_8$Ge$_4$ (5.12 mA/$\mu$m), or
Tl$_2$O (5.09 mA/$\mu$m), much less (17) as $p$-FET,
e.g. C$_2$N$_4$Pb$_2$ (4.08 mA/$\mu$m).

From the simulated ON-state characteristics we then extract the charge 
at the top-of-the-barrier (ToB) potential \cite{lundstrom},
$\rho_{ToB}$, and the carrier injection velocity at the same location,
$v_{inj}$=$I_{ON}$/($q\cdot\rho_{ToB}$), where $q$ is the elementary
charge (see Fig.~\ref{fig:2}d). These quantities are depicted in
Figs.~\ref{fig:2}e and \ref{fig:2}f, respectively. On average, the ToB
charge is equal to $\rho_{ToB,avg}$=1.55e13 cm$^{-2}$, with few
notable exceptions: several 2-D materials with a single $\Gamma$
electron valley, e.g. InX (with X=As, N, P, or Sb), Bi$_2$Se$_3$, or
Bi$_2$Te$_2$Se, are characterized by $\rho_{ToB}<$1e13
cm$^{-2}$. Their low effective mass $m^*\simeq$0.1 m$_0$, m$_0$ being
the electron rest mass, is responsible for that and for causing a
so-called density-of-states bottleneck effect: $\rho_{ToB}$ is no more
determined by the oxide capacitance of the dielectric layer, $C_{ox}$,
but limited by the quantum capacitance of the channel material, $C_Q$
\cite{dos}.   

In terms of injection velocities, all 2-D components with an ON-state
current larger than 4 mA/$\mu$m exceed the strained silicon marks of
both electrons (1.4e7 cm/s) and holes (1e7 cm/s) \cite{daa} and
therefore satisfy the IRDS objective. Some compounds in the ideal
corner of Fig.~\ref{fig:2}d, for example HfS$_2$, $n$-GeS, or $p$-SiH do
not fulfill this condition and have 10-20\% lower velocities than
expected. It can also be noticed in Fig.~\ref{fig:2}f that (i) up to
$I_{ON}$=2 mA/$\mu$m, the ON-state currents and injection velocities
are strongly correlated and (ii) the largest $v_{inj}$'s are found for
the 2-D materials with the lowest isotropic effective masses, which
are also those suffering the most from the DOS bottleneck.

These findings are the products of computationally intensive
simulations. We also aim to provide a more straightforward methodology
to predict the potential of 2-D materials without the need for complex
calculations. It is widely accepted that a semiconductor should have a
high concentration of fast carriers to deliver a large $I_{ON}$ at
fixed $I_{OFF}$. In the absence of source-to-drain tunneling, a low
transport ($m_{Trans}$) and a high density-of-states ($m_{DOS}$)
effective mass are required to supply these favorable conditions. In
2-D materials the $m_{Trans}$ and $m_{DOS}$ parameters cannot always
be unambiguously defined due to the presence of multiple band minima
(pockets), low energy separations between principal and satellite
valleys, and strong band non-parabolicities \cite{mounet}. Physically
meaningful averaging methods are needed to compute $m_{Trans}$ and
$m_{DOS}$. The proposed approach is described in the Methods Section.
It relies on the analytical expressions that can be derived for
single-valley semiconductors.

\begin{figure}[H]
\centering
\includegraphics[width=\linewidth]{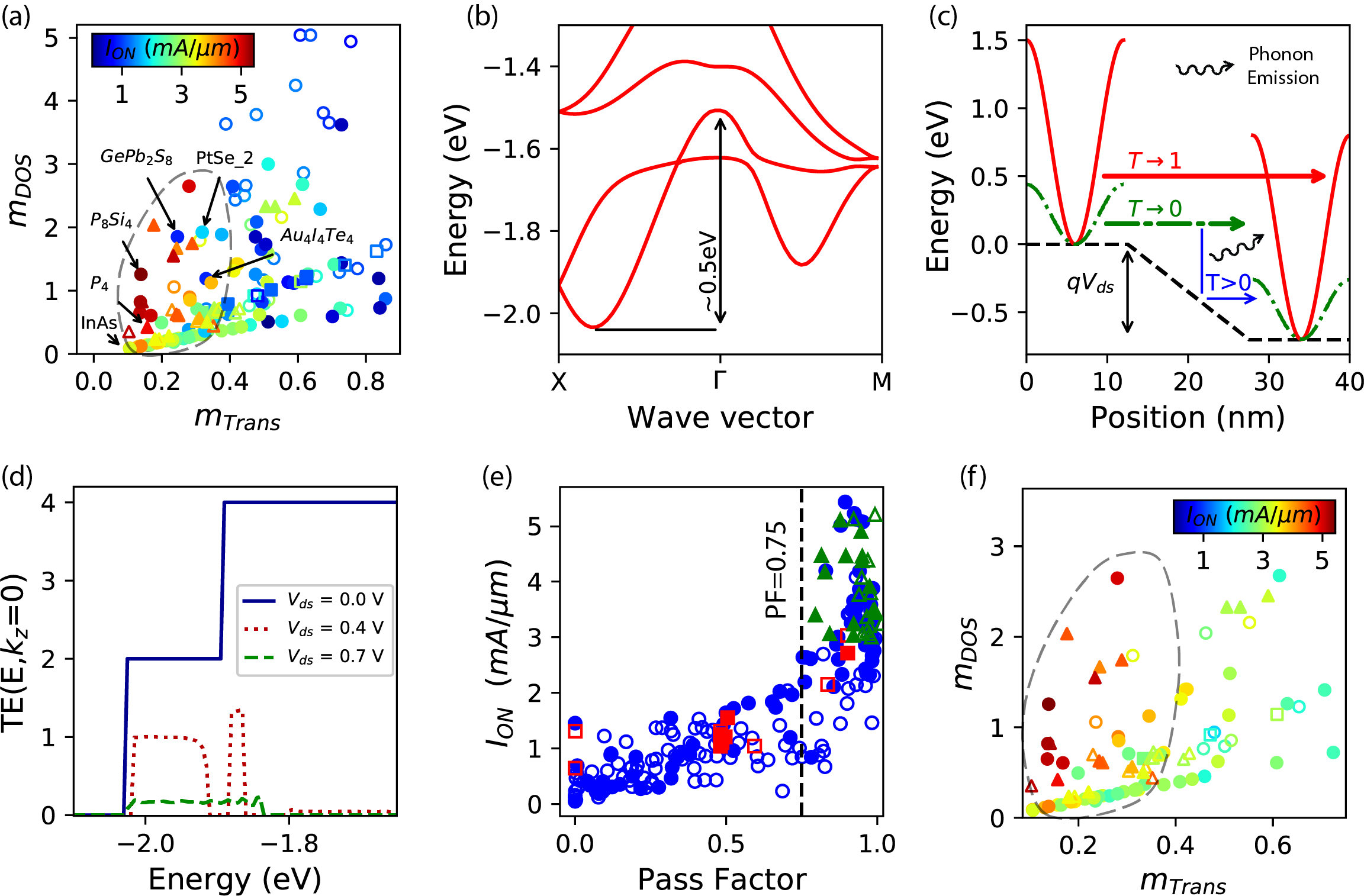}
\caption{
(a) ON-current of the 100 simulated $n$- and $p$-type 2-D FETs as a
  function of their transport $m_{Trans}$ and density-of-states $m_{DOS}$
  effective masses. Filled (empty) symbols refer to electron (hole)
  configurations, whereas squares represent TMDs and triangles the 13
  most promising 2-D materials. The color of each symbol indicates
  the magnitude of the corresponding $I_{ON}$. The shape in the lower
  left corner includes almost all best performing materials,
  i.e. those with a suitable ($m_{Trans}$,$m_{DOS}$) combination, but
  also several outliers.
(b) Bandstructure of AuITe around its conduction band minimum, along
  the X-$\Gamma$-M path. The lowest band has an energy width of
  $\sim$0.5 eV only, which is smaller than the applied source-to-drain
  voltage $V_{ds}$=0.7 V. 
(c) Schematic representation of the electron transmission through a
  device with an applied voltage $V_{ds}$. Only bands with an
  energy width larger than $qV_{ds}$ can directly carry current, those 
  with a narrower width must absorb or emit a phonon to be (partly) 
  transmitted.
(d) Energy-resolved transmission function $T(E,k_z=0)$ through a 40
  nm long AuITe monolayer with a linear potential drop of 0, 0.4, and
  0.7 V between both its extremities. Due to the presence of a narrow
  band, the transmission function strongly depends on the applied bias.
(e) Pass factor $PF$ of all considered 2-D FETs as a function of their
  ON-current. The same plotting conventions as in sub-plot (a)
  are used.
(f) Same as (a), but after eliminating the 2-D materials with
  $PF<$0.75. All components with a poor ON-state performance have
  disappeared from the shape in the lower left corner.
}
\label{fig:3}    
\end{figure}

Figure \ref{fig:3}a reports the ON-state currents of the 100 simulated
$n$- and $p$-type 2-D transistors as a function of their $m_{Trans}$ and
$m_{DOS}$ metrics. The best performing devices occupy the lower left
corner of the plot, as anticipated. They either possess an extremely
low $m_{Trans}$ and $m_{DOS}$, e.g. $n$-InAs, where $m_{Trans}\approx
m_{DOS}\approx$ 0.1$m_0$, or (moderately) low $m_{Trans}$ and
(relatively) large $m_{DOS}$ such as $n$-P$_8$Si$_4$
($m_{Trans}$=0.14$m_0$, $m_{DOS}$=1.26$m_0$) or $p$-P$_4$ 
with transport along the $\Gamma$-X axis ($m_{Trans}$=0.14$m_0$,
$m_{DOS}$=0.82$m_0$). While the shape in Fig.~\ref{fig:2}a encompasses
almost all FETs with $I_{ON}>$3 mA/$\mu$m, it also contains several
outliers that despite apparently suitable ($m_{Trans}$,$m_{DOS}$)
combinations return poor ON-state currents. This is the case of
$n$-PtSe$_2$ with transport in the $\Gamma$-K direction
($m_{Trans}$=0.32$m_0$, $m_{DOS}$=1.92$m_0$, $I_{ON}$=1.82
mA/$\mu$m) or $n$-Au$_4$I$_4$Te$_4$ ($m_{Trans}$=0.33$m_0$,
$m_{DOS}$=1.19$m_0$, $I_{ON}$=0.88 mA/$\mu$m).

The reason why $n$-PtSe$_2$ or $n$-Au$_4$I$_4$Te$_4$ show low
$I_{ON}$'s can be traced back to their bandstructure, which is
characterized at its bottom (conduction) or top (valence) by a single
band with an energy width smaller than the applied source-to-drain
voltage $V_{ds}$, as illustrated in Fig.~\ref{fig:3}b. In the
ballistic limit of transport such narrow bands contribute only partly
to the electrical current because a wave leaving the source contact
cannot find a matching state on the drain side (Fig.~\ref{fig:3}c)
\cite{szabo1}. It follows that the transmission function $T(E,k_z)$,
as computed with NEGF, is not only shifted by the gate-to-source
voltage, as in conventional FETs, but also changes its shape and
magnitude as the potential drop between source and drain increases
(Fig.~\ref{fig:3}d). The inclusion of electron-phonon scattering
reduces the negative impact of narrow bands by connecting them to
conducting channels through phonon emission and absorption
\cite{szabo1}. However, even under realistic conditions, i.e. in the
presence of dissipative scattering, narrow bands still limit the
transport properties of several 2-D materials. 

To quantify the influence of these narrow bands, we introduce a third
metric called the pass factor ($PF$). It describes the ratio between
the ballistic current that flows when a linear potential drop is
applied to a device, $I_{drop}$, and under flat band conditions,
$I_{flat}$, i.e. $PF=I_{drop}/I_{flat}$ (see Methods Section). In
$I_{drop}$ the current contributions coming from the narrow bands are
filtered out, whereas they are accounted for in $I_{flat}$. It is
important to note that the smaller the $PF$ is, the more
perturbed by narrow bands is the propagation of electrons and holes
through 2-D devices. When $PF$ is 1, transport becomes optimal. The
pass factor of the considered $n$- and $p$-type FETs is depicted in
Fig.~\ref{fig:3}e. It can be seen that all 2-D materials with a large
ON-current have a $PF$ greater than 0.75. To reflect this observation
we plot again in Fig.~\ref{fig:3}f the ``$I_{ON}$ vs. $m_{Trans}$ and
$m_{DOS}$'' data from before, this time keeping only the components
with $PF>$0.75. All outliers in the lower left corner of
Fig.~\ref{fig:3}a have now disappeared, demonstrating that besides the
transport ($m_{Trans}$) and density-of-states ($m_{DOS}$) effective
masses, a third parameter, the pass factor ($PF$), is necessary to
reveal the most promising candidates as ultimate 2-D logic switches
and remove ``false positive'' contenders.  

\textbf{Gate Length Scalability.} After studying the performance of
2-D FETs at $L_g$=15 nm, we turn to their scalability down to 5 nm. To
bring insight into this critical issue, the subthreshold swing (SS) of
all simulated transistors is extracted at different gate lengths (15,
12.5, 10, 7.5, and 5 nm) over a voltage range $\Delta V_{gs}$=0.1 V at
$V_{ds}$=0.7 V. The results are reported in the form of a histogram in
Fig.~\ref{fig:4}a, whereas the median ``SS vs. $L_g$'' data averaged
over all 2-D $n$- and $p$-FETs is presented in Fig.~\ref{fig:4}b. Down
to a gate length of 10 nm, the SS distribution is very localized, with
average values of 72 mV/dec at $L_g$=15 nm, 77 mV/dec at $L_g$=12.5
nm, and 88 mV/dec at $L_g$=10 nm, with relatively narrow standard
deviations of 3.4, 11.9, and 15.1 mV/dec, respectively. Starting at
$L_g$=7.5 nm, the SS spread drastically increases, indicating that
single-gate 2-D FETs might not work optimally below 10 nm gate
lengths. It should however be emphasized that with SS$_{avg}$=72
mV/dec at $L_g$=15 nm and SS$_{avg}$=77 mV/dec at $L_g$=12.5 nm, these
devices surpass the 2025 IRDS targets for multi-gate Si FinFETs (80
mV/dec) \cite{irds}. 

As already mentioned, 2-D materials with a high $I_{ON}$ typically
have a low transport effective mass $m_{Trans}$, which can lead to
large source-to-drain (S-to-D) tunneling leakages and poor
sub-threshold swings at short gate lengths. No such correlation is
noticed in Fig.~\ref{fig:4}c, where SS at $L_g$=15 nm is reported as a
function of the transistor ON-currents. It can thus be inferred that
S-to-D tunneling plays a minor role at this gate length for the
2-D semiconductors considered and that the calculated band gaps, even
if underestimated by DFT, do not impact SS at this scale, except for
Bi$_2$GeTe$_4$ and InSb. Their narrow $E_g$ of 0.99 and 1.07 eV,
respectively, combined with a low $m_{Trans}$=0.11$m_0$ for
electrons in InSb, push SS above 80 mV/dec. Note that at gate lengths
$L_g$=7.5 and 5 nm, when S-to-D tunneling starts to severely limit SS,
$E_g$ becomes a crucial parameter. Hence, from a quantitative point of
view, the simulation results below $L_g$=10 nm should be taken with
precautions as the present DFT+NEGF approach may overestimate the
sub-threshold swing of certain 2-D materials, especially those with a
DFT band gap smaller than 1.2 eV and a low $m_{Trans}$. It remains
that even for compounds with band gaps larger than 1.5 eV, the average
SS goes beyond 100 mV/dec at $L_g$=5 nm, which calls for alternative
device designs.

The scalability of transistors down to $L_g$=5 nm can be improved by
using gate-underlapping, i.e. by keeping undoped regions around the
gate contact. This technique allows to increase the effective gate
length without modifying its physical footprint \cite{iedm11}. Lower
ON-currents usually result from this process. Alternatively, the
bottom SiO$_2$ substrate in Fig.~\ref{fig:1}a can be replaced by a
thin HfO$_2$ layer of the same thickness as the top one ($t_{ox}$=3
nm) and a second gate electrode can be inserted below it, symmetric
with the first one. Such double-gate (DG) architectures, as
illustrated in Fig.~\ref{fig:4}d are more complex and expensive to
fabricate than single-gate ones, but they substantially reduce SS over
the whole gate length range of interest, as demonstrated in the SI for
selected 2-D materials. This phenomenon can be explained by examining
the dependence of SS on the oxide capacitance $C_{ox}$, when S-to-D
tunneling is ignored. It is given by SS=(1+$C_Q$/$C_{ox}$)$\times$60
mV/dec, at room temperature \cite{rahman}, where $C_Q$ is the quantum
capacitance of the channel material. By adding a second gate contact
acting in parallel with the first one, $C_{ox}$ increases by a factor
up to 2, the term 1+$C_Q$/$C_{ox}$ gets closer to 1, and SS approaches
60 mV/dec.

Double-gate FETs do not only scale better than their single-gate
counterparts; they also show higher ON-currents since their SS is 
steeper than in SG configurations (more rapid increase of the current)
and their $C_{ox}$ is larger (more mobile charges are generated
at the top-of-the-barrier location). The combined influence of these
effects can more than double the ON-state current of DG transistors, as
compared to SG ones, as can be seen in Figs.~\ref{fig:4}e and
\ref{fig:4}f for a GeS $n$- and $p$-type FET. Results for other 2-D
materials with high ON-currents are presented in Table \ref{tab:2}. As
expected, in all cases, the DG $I_{ON}$ is between 1.9 and 2.3 times
larger than the SG one, attaining values above 10 mA/$\mu$m in few
materials (As$_4$, P$_4$, and As$_8$Ge$_4$). If such devices could be
successfully fabricated and their contact resistance kept below 100
$\Omega\cdot\mu$m, they would outplay the IRDS projections for Si
FinFETs by a factor 6. More DG results can be found in the SI. 

To complete the analysis of Table \ref{tab:2} it should be underlined
that $\rho_{ToB}$ does not increase by the same amount as the current
(1.6$\times$ for $\rho_{ToB}$ vs. 1.9 to 2.3$\times$ for $I_{ON}$)
when going from single- to double-gate transistors. From this
observation it can be deduced that the injection velocity at the 
ToB, $v_{inj}$, is greater in the DG structures too. This does not
come as a surprise since higher-energy states get populated when
$\rho_{ToB}$ increases. These states are usually characterized by a
steeper band dispersion than those situated close to a band extremum
and therefore by a higher injection velocity, 30\% on average for the
considered logic devices.

\section{Conclusions}
We performed \textit{ab initio} quantum transport simulations of 100
different single-gate $n$- and $p$-type field-effect transistors using
stable 2-D monolayers taken from the computational database of
Ref.~\cite{mounet} as channel materials. For each of them, the
\textit{I-V} characteristics at a gate length $L_g$=15 nm were
calculated and their scalability was investigated down to $L_g$=5
nm. The ON-current of 13 2-D materials was found to be larger than 3
mA/$\mu$m in both their $n$- and $p$-type configuration, even
exceeding 4 mA/$\mu$m in three cases. Excellent sub-threshold swings
down to $L_g$=10 nm were obtained, confirming the great scalability
potential of the 2-D technology. Finally, the compounds delivering the
highest ON-currents were further tested with a double-gate
architecture, which increased their $I_{ON}$ by a factor 2 and might
be required to go below $L_g$=10 nm. These findings clearly indicate
that 2-D materials could enable the continuation of Moore's scaling
law beyond Si FinFETs, provided that their contact resistances improve
and their structural as well as interfacial defect density is
reduced. Our results should be seen as the starting point for the
exploration of novel 2-D materials with better transport properties
than the known ones. We hope that they will motivate experimental
groups to exfoliate new compounds, e.g. As$_8$Ge$_4$, O$_6$Sb$_4$, or
C$_2$N$_4$Pb$_2$ that are expected to outperform the popular
transition metal dichalcogenides and challenge Si-based transistors.  

On the simulation side, the design space will be extended with 2-D
materials coming from other databases \cite{jarvis,den} and from 
on-going searches conducted by some of the co-authors of this
study. The proposed metrics ($m_{Trans}$, $m_{DOS}$, and $PF$) will be
extracted first. This will limit the number of 2-D systems that must
be simulated at the quantum transport level and reduce the
computational burden. The current 13 best-performing 2-D materials
will go through a second round of analysis where electron-phonon,
defect, and surface optical phonon scattering will be added to
accurately model their carrier mobility \cite{iedm19}. The dielectric 
environment surrounding the 2-D monolayers will be explicitly
accounted for in the Schr\"odinger equation; when possible, GW
corrections and spin-orbit coupling will be included as well. Contact
resistances will also be evaluated with different metal
electrodes. Altogether this will help refine the predictions made in
this paper.

\section{Methods}
\noindent\textbf{Selection of 2-D Materials.}
The database of Ref.~\cite{mounet} containing more than 1,800 entries
was pre-screened to select 100 2-D materials that were then
investigated as single-gate planar transistors. Based on their 
bandstructure characteristics we singled out the semiconductors with
less than 30 atoms in their primitive unit cell, a layer thickness
smaller than 1.5 nm, a band gap larger than 1 eV and, if possible,
anisotropic conduction band minima and/or valence band maxima so that
a low transport and high density-of-states effective mass can be
obtained. We also eliminated 2-D materials where the lowest conduction
or highest valence subband had a very narrow energy width of 0.5 eV or 
less and was completely isolated from the other bands. Finally, we
inspected the structural stability of the chosen 2-D components by 
verifying that their phonon bandstructure did not exhibit negative
branches.  

\noindent\textbf{DFT Calculations.}
The first step of each DFT calculation consists of identifying the
primitive unit cell of the considered 2-D material. The lattice
vectors and atomic positions are taken from Ref.~\cite{mounet}. A
vacuum space of 20 \AA\phantom{ }is added along the confined direction
to remove fictitious interactions between a 2-D layer and its periodic
image. For the $k$-point sampling of the Brillouin Zone
$\Gamma$-centered Monkhorst-Pack grids are employed with equal
densities of $k$-points in all directions, except for the confined
one. The plane-wave cutoff is set to at least 1.5 times the 
default value found in the exchange-correlation potential. For
computational reasons spin-orbit coupling is neglected. Its influence
on the transport properties of the simulated 2-D materials can
therefore neither be quantified nor estimated. 

Most electronic structures were calculated within the generalized
gradient approximation (GGA) of Perdew, Burke, and Ernzerhof
(PBE)\cite{pbe} with the DFT-D3 parameterization of van der Waals
forces from Grimme\cite{grimme}. At the beginning of this work, few
2-D materials were treated with the optPBE-vdW functional of Klimes,
Bowler, and Michaelides \cite{klimes} before realizing that this model
sometimes leads to the presence of valence states in the vacuum region
of the simulation domain. For this reason optPBE-vdW  was later on
abandoned and 2-D materials whose electronic structure was obtained
with it are specifically marked in the SI. Note that it has been verified
that in case of perfectly empty vacuum, optPBE-vdW and PBE give the
same bandstructure results.

For each 2-D sample, three consecutive DFT calculations are performed
with the Vienna Ab initio Software Package (VASP) \cite{vasp}. First,
the atomic positions within the unit cell are relaxed until the force
acting on each ion is converged to below 10$^{-3}$ eV/\AA. The
electronic structure is determined in a second step based on a very
fine $k$-point grid so that the final energies vary by less than 1 meV
per atom when increasing the number of $k$-points. Finally, the
bandstructure along a path through the high-symmetry points of the
Brillouin Zone is calculated with the charge density from the
second simulation. 

Although DFT is an \textit{ab initio} method its results may depend
on the choice of the exchange-correlation functional, basis set, or
pseudo-potential parameterization. As a ground-state theory it tends
to underestimate the band gap of semiconductors, which might lead
to inaccurate predictions when source-to-drain tunneling becomes
important, at gate lengths of 10 nm and below. GW corrections
\cite{louie} usually improve the situation, but with 2-D materials,
this is not necessarily the case as their band gaps, effective masses,
and valley splitting may depend on the surrounding dielectric
environment \cite{kharche}. As such they are currently not known
exactly, preventing a validation of the obtained bandstructures.
Despite these drawbacks, in the vast majority of the cases, the fact
that the precise band gap value remains uncertain does not impact the 
presented results, especially when $L_g>$10 nm. This is why the
simulation data proposed in this paper give a unique overview on the
functionality of 100 different 2-D FETs.

\begin{figure}[H]
\centering
\includegraphics[width=0.85\linewidth]{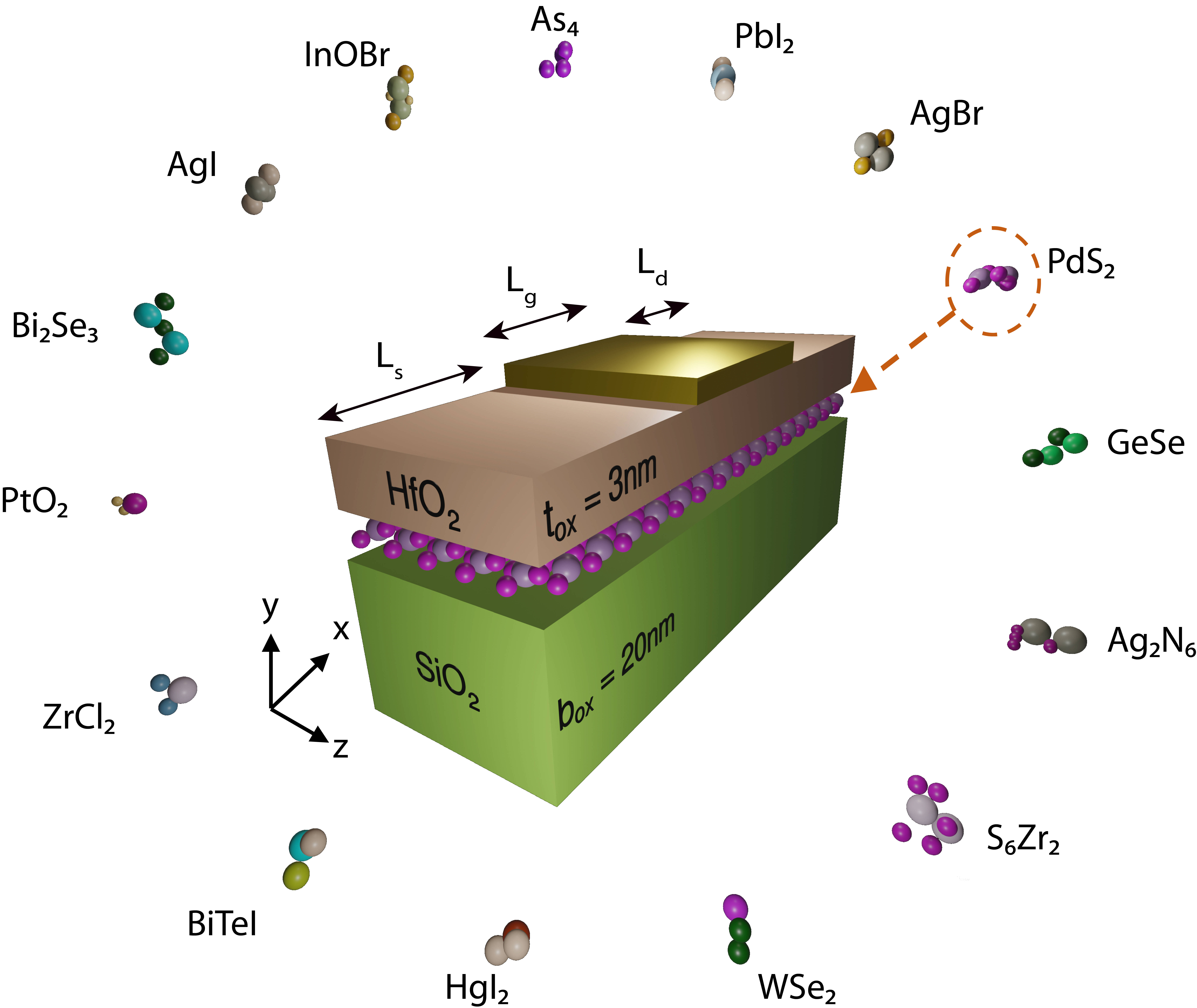}
\caption{
  Schematic view of the single-gate field-effect transistors (SG
  FETs) investigated in this work. The total device length measures 40
  nm, with a gate length $L_g$ varying between 15 and 5 nm. The source
  and drain extensions are doped with a donor $N_D$ or acceptor $N_A$
  concentration of 5e13 cm$^{-2}$ for the $n$- and $p$-type configuration,
  respectively. The doping abruptly stops at the channel-source/drain
  interfaces. A $t_{ox}$=3 nm thick high-$\kappa$ dielectric layer 
  (HfO$_2$) separates the 2-D channel from the gate contact. The 2-D
  material (here PdS$_2$) is deposited on a SiO$_2$ box of thickness
  $t_{box}$=20 nm. Perfectly ohmic contacts are assumed (no
  resistance). All simulations are performed at room temperature with
  the metal gate work function adjusted to fix the OFF-current to 0.1
  $\mu$A/$\mu$m. Transport occurs along the $x$-axis, $y$ is a
  direction of confinement, while $z$ is assumed periodic. The
  primitive unit cells of few representative 2-D materials are plotted
  around the transistor structure.}
\label{fig:1}    
\end{figure}

\begin{figure}[H]
\centering
\includegraphics[width=\linewidth]{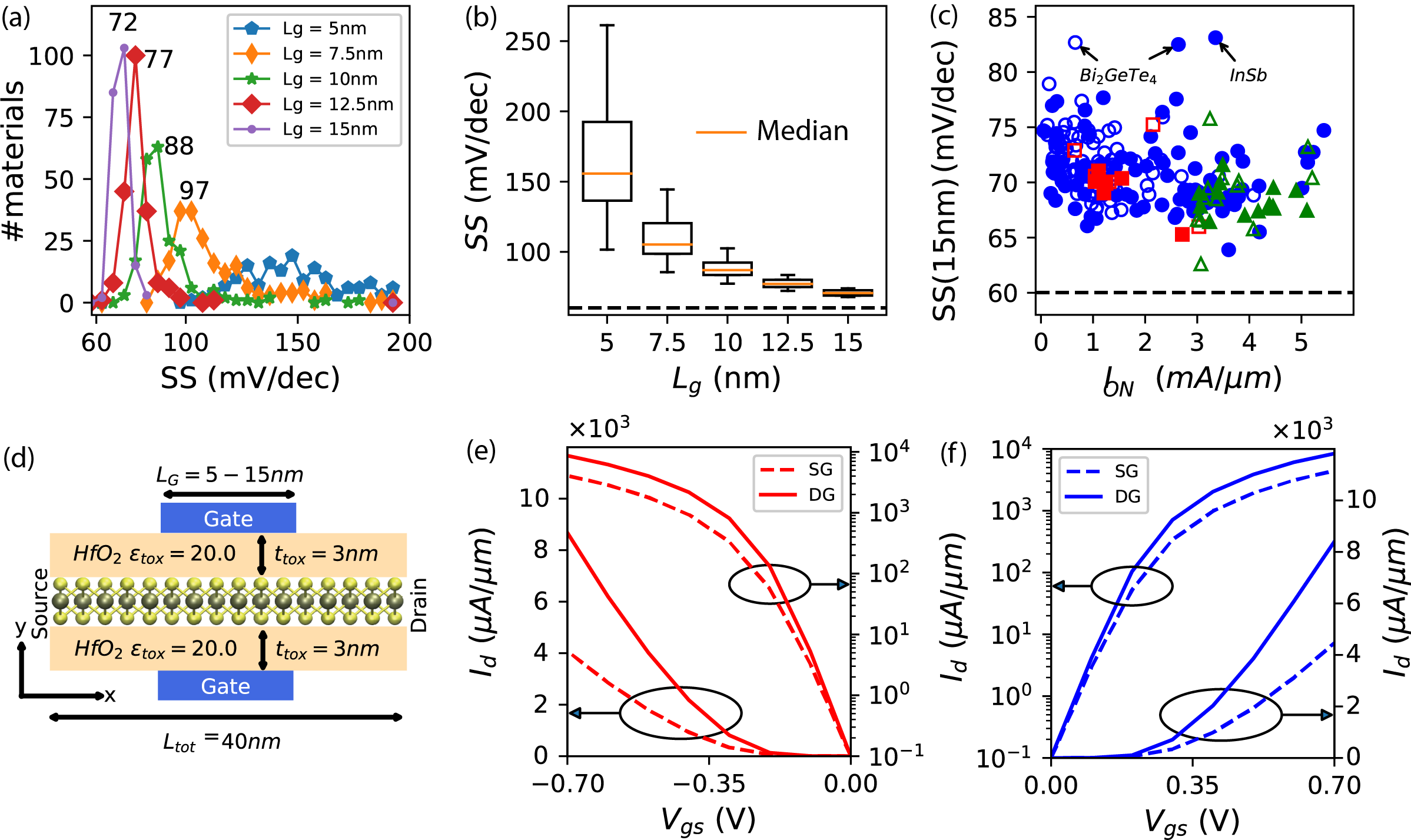}
\caption{
(a) Histogram of the sub-threshold swing (SS) of all investigated $n$-
  and $p$-type FETs at a gate length $L_g$=15, 12.5, 10, 7.5, and 5
  nm. The mean value is indicated for each $L_g>$5 nm. 
(b) Box plot summarizing the distribution of the sub-threshold swings
  for the considered gate lengths. Each box contains 50\% of the
  extracted SS. The mark in the middle of the box is the median. The
  ends of the whiskers represent one standard deviation above and
  below the mean of the data. 
(c) Sub-threshold swing at $L_g$=15 nm as a function of the
  ON-current at the same gate length. Green triangles refer to the 13
  best 2-D materials, red squares to conventional TMDs, while filled
  (empty) symbols represent $n$-type ($p$-type) transistors.
(d) Schematic view of a double-gate GeS FET. As compared to the
  structure in Fig.~\ref{fig:1}, the bottom SiO$_2$ substrate was
  replaced by a HfO$_2$ layer of thickness $t_{ox}$=3 nm and a second
  gate contact was added at the bottom.
(e) Transfer characteristics $I_d$-$V_{gs}$ at $V_{ds}$=0.7 V of the
  GeS $p$-FET in (d) with $L_g$=15 nm, using a single- and double-gate
  architecture. The \textit{I-V} curves are plotted on a linear and
  logarithmic scale.
(f) Same as (e), but for the GeS $n$-FET.  
}
\label{fig:4}    
\end{figure}

\noindent\textbf{Wannier Functions.}
The plane-wave results produced by VASP are transformed into a basis
of maximally-localized Wannier functions (MLWFs) with the wannier90
\cite{wannier90} tool. An energy window encompassing 0.7 to 1 eV
around the band gap of each material is defined. All bands originating
from within this interval are retained in the MLWF Hamiltonian. Next
the latter is diagonalized for the same $k$-point path as the
original VASP simulation. The resulting VASP and MLWF bandstructures
can then be compared to each other. The Wannierization process is
considered as successful if all occupied bands, according to the
chosen doping concentration (5e13 cm$^{-2}$), are reproduced within a
maximum tolerance of 20 meV and if the errors around the conduction and
valence band edges are below 5 meV. As last step the MLWF
Hamiltonian blocks corresponding to the primitive unit cell of the
studied 2-D materials are scaled up according to the techniques
presented in Ref.~\cite{szabo2} to yield the Hamiltonian matrix of the
full device, $H_{MLWF}$ (see Fig.~\ref{fig:1}a for the dimensions). 

\noindent\textbf{Quantum Transport Simulations.}
The OMEN code is used as quantum transport simulator \cite{omen}. The
approach is very similar to the one described in the Supporting
Information of Ref.~\cite{szabo2}. After upscaling the MLWF
Hamiltonian the \textit{I-V} characteristics of the investigated 2-D
transistors are computed with the NEGF formalism at room
temperature. The retarded ($G^R(E,k_z)$), lesser ($G^<(E,k_z)$), 
and greater (($G^>(E,k_z)$)) Green's function are evaluated for all
possible carrier energies $E$ and momentum points $k_z$ using
 \begin{eqnarray} 
\left(E-H_{MLWF}(k_z)-\Sigma^{RB}(k_z,E)-\Sigma^{RS}(k_z,E)\right)\cdot
G^R(k_z,E)\quad=\quad I,\label{eq:a}\\
G^{\gtrless}(k_z,E)\quad=\quad G^R(k_z,E)\cdot\left(\Sigma^{\gtrless
  B}(k_z,E)+\Sigma^{\gtrless S}(k_z,E)\right)\cdot G^A(k_z,E).
\label{eq:b}
\end{eqnarray}
The advanced Green's Function $G^A(E,k_z)$ is the transpose of
$G^R(E,k_z)$. The Hamiltonian matrix $H_{MLWF}$ is expressed in a MLWF
basis and constructed according to the prescriptions described
above. The size of $H_{MLWF}$ as well as of all Green's Functions is
equal to $N_A\times N_{wf}$, where $N_A$ is the total number of atoms
in the simulated system and $N_{wf}$ the average number of Wannier
functions per atom.

In Eqs.~(\ref{eq:a}) and (\ref{eq:b}), the energy vector is
homogeneously discretized with a distance $dE$=1 meV between two
adjacent points. The momentum dependence, which is a consequence of
the assumed-periodic out-of-plane direction $z$ in Fig.~\ref{fig:1}a, is
modeled via $N_{kz}$=15 points. Orthorhombic unit cells are used in our 
transport calculations. Their width along the $z$ direction is chosen
to be larger than 1 nm so that $N_{kz}$=15 is sufficient to obtain
accurate results. The open boundary conditions and all
scattering sources are cast into the self-energies $\Sigma^{TB}(E,k_z)$
and $\Sigma^{TS}(E,k_z)$, respectively. They are of the same type $T$  
($R$, $<$, or $>$) and size as the Green's Functions.
  
Non-physical behaviors may occur in 2-D materials in the ballistic
limit of transport due to the presence of narrow bands
\cite{szabo1}. To avoid them and provide realistic simulation results,
an energy relaxing mechanism is introduced into the NEGF equations. We
opted for a phenomenological electron-phonon scattering model where
two parameters can be freely chosen, the phonon frequency $\omega$ and
the deformation potential $D_{e-ph,}$ 
\begin{eqnarray}
\Sigma^{\lessgtr S}(k_z,E)&=&D^2_{e-ph}\left(n_{\omega}G^{\lessgtr}(k_z,E+\hbar\omega)+(n_{\omega}+1)G^{\lessgtr}(k_z,E-\hbar\omega)\right). 
\label{eq:c}
\end{eqnarray}
In Eq.~(\ref{eq:c}) $n_{\omega}$ is the Bose-Einstein distribution
function. We use a phonon energy $\hbar\omega$=40 meV and a scattering
strength $D_{e-ph}$ comprised between 25 and 125 meV, depending on the
width of the narrow bands. The resulting equations are solved within
the so-called self-consistent Born approximation till convergence is
reached, i.e. till the electronic current is conserved all along the
device structure. By considering these (pseudo) electron-phonon
interactions, energy bands with a narrow width get connected to each
other, either through phonon emission or absorption, which opens
additional conduction channels and eliminates ballistic modeling
artifacts. Furthermore, non-idealities such as inter-valley scattering
are accounted for to a certain extent. Still, all investigated 2-D
transistors operate close to their  ballistic limit: the used
scattering strength causes a decrease of the electrical current by no
more than 20\%. Note however that for the 2-D materials with narrow
energy bands, an increase of the current is typically observed when
dissipative interactions are turned on \cite{szabo1}.  

\noindent\textbf{Effective Mass and Pass Factor Extraction.}
To extract the transport ($m_{Trans}$) and density-of-states
($m_{DOS}$) effective masses of 2-D materials and allow for meaningful
comparisons of different components, the average ToB charge density
$\rho_{ToB,avg}$=1.55e13 cm$^{-2}$ from Fig.~\ref{fig:1}e is
recalled. All calculations are done assuming that the bandstructure of
the 2-D materials is filled with that specific density, which sets a
reference to establish the potential of a crystal before running any
device simulation.  

First, the bandstructure $E(k_x,k_z)$ of the 2-D semiconductor under
investigation is computed for all ($k_x$,$k_z$) wave vectors belonging
to its Brillouin Zone (BZ), see the SI. Here, $k_x$ refers to the axis
aligned with the transport direction, $k_z$ models the one assumed
periodic, i.e. the out-of-plane direction in Fig.~\ref{fig:1}a. In
this study, an orthorhombic unit cell is used as starting point. With
the knowledge of $E(k_x,k_z)$ the Fermi level $E_f$ that corresponds
to $\rho_{ToB,avg}$ can be numerically acquired from 
\begin{eqnarray}
\rho_{ToB,avg}&=&\frac{1}{A}\sum_{k_x,k_z\in BZ}f(E(k_x,k_z),E_f),
\label{eq:1}
\end{eqnarray}
where $A$ is the area of the unit cell and $f(E,E_f)$ the Fermi-Dirac
distribution function. The sum goes over the entire Brillouin Zone,
which is rectangular for orthorhombic unit cells. In case of holes,
$E(k_x,k_z)$ is first multiplied by -1 so that all equations derived
for electrons can be directly applied. From now on, no distinction is
made in the treatment of electrons and holes.     

Next, the bandstructure-limited ToB electronic current $I_{d,BS}$ that
flows through an idealized device at a charge concentration
$\rho_{ToB,avg}$=1.55e13 cm$^{-2}$ is evaluated with the following 
equation
\begin{eqnarray}
I_{d,BS}&=&\frac{1}{2A}\sum_{k_x,k_z\in BZ}\left|v_x(k_x,k_z)\right|f(E(k_x,k_z),E_f)
\label{eq:2}
\end{eqnarray}
with the $E_f$ value from Eq.~(\ref{eq:1}). The factor $1/2$ indicates
that only half of the Brillouin Zone carries current, the one where
the electronic states have a positive velocity
$v_x(k_x,k_z)$=$1/\hbar\phantom{.} dE(k_x,k_z)/dk_x$ along the 
transport axis $x$ in Fig.~\ref{fig:1}a. The variable $\hbar$ is
Planck's reduced constant.

If a single parabolic band with an anisotropic dispersion
$E(k_x,k_z)$=$E_{min}$+$\hbar^2/2(k_x^2/m_x+k_z^2/m_z)$ and an energy
minimum $E_{min}$ is assumed, then quasi-analytical expressions can be
found for the solution of Eqs.~(\ref{eq:1}) and (\ref{eq:2}) 
\begin{eqnarray}
\rho_{ToB,avg}&=&\frac{\sqrt{m_xm_z}}{\pi\hbar^2}\int_{E_{min}}^{\infty}dE\phantom{.}f(E,E_f),\label{eq:3}\\
I_{d,BS}&=&\frac{q\sqrt{2m_z}}{\pi^2\hbar^2}\int_{E_{min}}^{\infty}dE\phantom{.}\sqrt{E-E_{min}}\phantom{.}f(E,E_f).\label{eq:4}
\end{eqnarray}
We observe that $m_{DOS}$=$\sqrt{m_xm_z}$ and by definition,
$m_{Trans}$=$m_x$, the effective mass along the transport
direction. For comparison purpose, the bandstructure of all considered
2-D materials is reduced to these two highly relevant
quantities that can be interpreted as occupancy-aware, averaged
masses. First, $E_f$ is computed in Eq.~(\ref{eq:1}), then 
$I_{d,BS}$ in Eq.~(\ref{eq:2}), $m_{DOS}$ in Eq.~(\ref{eq:3}), and
finally $m_{Trans}$=$m_{DOS}^2/m_z$ in Eq.~(\ref{eq:4}). The obtained
effective masses indirectly account for all valleys situated close to
the band extrema, non-parabolic effects, and band filling according to
the average ON-state charge at the ToB location, $\rho_{ToB,avg}$. No
QT simulation is needed to evaluate Eqs.~(\ref{eq:1}) to (\ref{eq:4}),
only bandstructures. 

\begin{figure}[H]
\centering
\includegraphics[width=\linewidth]{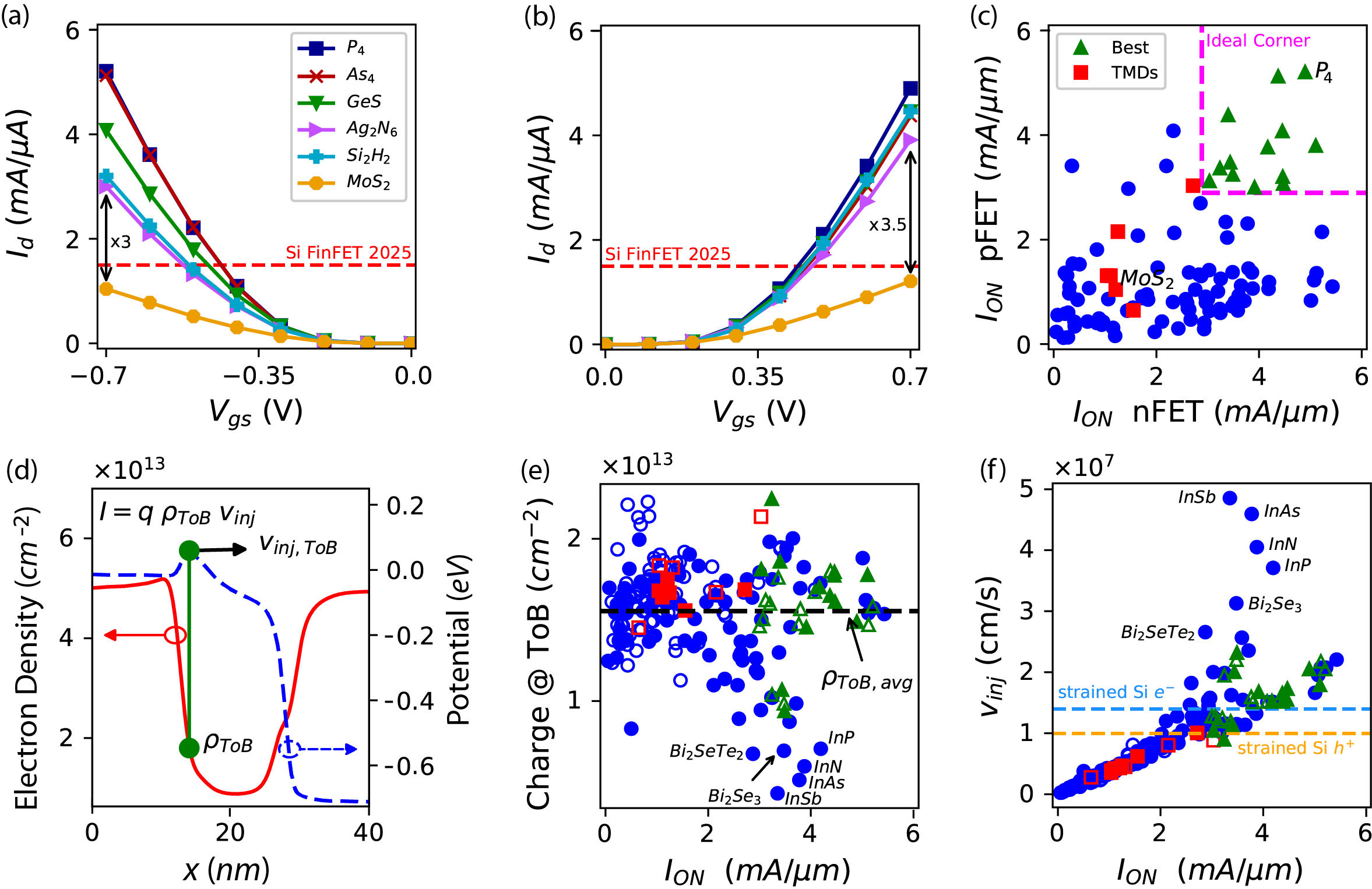}
\caption{
(a) Transfer characteristics $I_d$-$V_{gs}$ at $V_{ds}$=$V_{DD}$=0.7 V
  of 6 selected 2-D $p$-type FETs similar to the one in
  Fig.~\ref{fig:1}: P$_4$, As$_4$, GeS, Ag$_2$N$_6$, SiH, and MoS$_2$.
(b) Same as (a), but for the $n$-type configuration.
(c) ``$n$-type $I_{ON}$ vs. $p$-type $I_{ON}$'' for the 100 simulated 2-D
  materials. Green triangles refer to components with both current
  types larger than 3 mA/$\mu$m, red squares to standard TMDs (MX$_2$
  with M=Mo or W and X=S, Se, or Te).
(d) Illustration of the method used to extract the top-of-the-barrier
  location \cite{lundstrom} from 2-D devices as well as the
  corresponding charge $\rho_{ToB,avg}$ and injection velocity $v_{inj}$.   
(e) ON-state charge density extracted at the top-of-the-barrier
  location as a function of the ON-current. The dashed line indicates
  the average charge over all devices, $\rho_{ToB,avg}$=1.55e13
  cm$^{-2}$. Filled (empty) symbols correspond to electron (hole) conduction.
(f) ON-state carrier injection velocity $v_{inj}$ at the ToB location
  with respect to the ON-current. The values for electrons and holes
  in strained silicon are given as references \cite{daa}.
}
\label{fig:2}    
\end{figure}

The last metrics that we used to assess the potential of a given 2-D
material is a new quantity called pass factor ($PF$). It is defined as
\begin{eqnarray}
PF&=&\frac{\sum_{k_z}\int_{E_{min}}^{\infty}dE\phantom{.}T_{V_{ds}=V_{DD}}(E,k_z)f(E,E_f)}
{\sum_{k_z}\int_{E_{min}}^{\infty}dE\phantom{.}T_{V_{ds}=0}(E,k_z)f(E,E_f)},
\end{eqnarray}
where $T_{V_{ds}}(E,k_z)$ is the energy- and momentum-dependent
transmission function calculated when a linear potential drop of
amplitude $qV_{ds}$ is applied between both ends of a 2-D material, as
in Fig.~\ref{fig:2}d. Note that determining $T_{V_{ds}}(E,k_z)$
requires a non-self-consistent QT simulation involving $N_E$=1,000
energy- and $N_{k_z}$=15 momentum-points, which remains
computationally very affordable. 

The $m_{Trans}$, $m_{DOS}$, and $PF$ values of the simulated 2-D
materials can be found in the SI.

\begin{acknowledgement}
This research was supported by ETH Zurich (grant ETH-32 15-1) and
by the Swiss National Science Foundation (SNSF) under grant
no. 200021\_175479 (ABIME) and under the NCCR MARVEL. We acknowledge
PRACE for awarding us access to Piz Daint at CSCS under Project pr28,
PRACE for the allocated computational resources on Marconi at CINECA
under Project 2016163963, and CSCS for Project s876.
\end{acknowledgement}

\begin{suppinfo}
Electron/hole ``current vs. voltage'' characteristics, ``sub-threshold
swing vs. gate length'', unit cell, bandstructure, injection velocity,
transport and density-of-states effective masses, and ``pass factor''
of the 100 2-D materials considered in this work.
\end{suppinfo}

\newpage

\newpage

\newpage

 \newpage

 \begin{table}
{\footnotesize
\begin{tabular}{|l|cccc|cccc|}
\cline{2-9}
\multicolumn{1}{c|}{} & \multicolumn{4}{c|}{$n$-FET} & \multicolumn{4}{c|}{$p$-FET}\\
\cline{2-9}         
\multicolumn{1}{c|}{} & \makecell{$I_{ON}$\\(mA/$\mu$m)} & \makecell{$\rho_{ToB}$\\(cm$^{-2}$)} 
& \makecell{$v_{inj}$\\(cm/s)}  & \makecell{SS\\(mV/dec)} &
\makecell{$I_{ON}$\\(mA/$\mu$m)} & \makecell{$\rho_{ToB}$\\(cm$^{-2}$)} 
& \makecell{$v_{inj}$\\(cm/s)} & \makecell{SS\\(mV/dec)}  \\
\hline
  Ag$_2$N$_6$       &   3.91  &  1.45e13 &  1.68e7  &  67  &  3  &  1.46e13  &  1.28e7  &  66.6  \\
\hline
  As$_2$            &   4.17  &  1.7e13 &  1.53e7  &  67.4  &  3.77  &  1.47e13  &  1.59e7  &  69.7  \\
\hline
  As$_4$            &   4.37  &  1.65e13 &  1.65e7  &  67.7  &  5.12  &  1.47e13  &  2.18e7  &  73.2  \\
\hline
  Ge$_2$S$_2$       &   4.45  &  1.77e13 &  1.57e7  &  67.8  &  4.08  &  1.67e13  &  1.52e7  &  65.8  \\
\hline
  Ge$_2$Se$_2$      &   3.4  &  1.86e13 &  1.14e7  &  69.1  &  4.38  &  1.8e13  &  1.52e7  &  68  \\
\hline
  HfS$_2$           &   3.27  &  1.97e13 &  1.04e7  &  69.1  &  3.59  &  1.77e13  &  1.26e7  &  68.8  \\
\hline
  O$_6$Sb$_4$       &   4.47  &  1.61e13 &  1.73e7  &  67.6  &  3.07  &  1.51e13  &  1.27e7  &  62.6  \\
\hline
  P$_4$             &   4.9  &  1.49e13 &  2.05e7  &  69.3  &  5.21  &  1.58e13  &  2.05e7  &  70.4  \\
\hline
  Sb$_2$            &   5.11  &  1.78e13 &  1.79e7  &  67.5  &  3.8  &  1.57e13  &  1.51e7  &  70.1  \\
\hline
  Si$_2$H$_2$       &   4.46  &  1.8e13 &  1.55e7  &  69.5  &  3.2  &  1.6e13  &  1.25e7  &  69.5  \\
\hline
  Ti$_2$Br$_2$N$_2$ &   3.49  &  9.38e12 &  2.32e7  &  71.6  &  3.25  &  1.04e13  &  1.95e7  &  75.8  \\
\hline
  Ti$_2$N$_2$Cl$_2$ &   3.44  &  1.07e13 &  2.01e7  &  69.9  &  3.48  &  9.85e12  &  2.2e7  &  70.4  \\
\hline
  ZrS$_2$           &   3.24  &  2.25e13 &  9e6  &  66.4  &  3.37  &  1.76e13  &  1.19e7  &  68.5  \\
\hline
\end{tabular}
}
\caption{ON-state current ($I_{ON}$), charge at the top-of-the-barrier
  ($\rho_{ToB}$), injection velocity at the same location ($v_{inj}$),
  and sub-threshold swing (SS) for the 13 2-D single-gate FETs with
  $I_{ON}>$3 mA/$\mu$m in both their $n$- and $p$-type configuration
  at $L_g$=15 nm.}
\label{tab:1}
\end{table}

\newpage

\begin{table}
{\footnotesize
\begin{tabular}{|l|cccc|cccc|}
\cline{2-9}
\multicolumn{1}{c|}{} & \multicolumn{4}{c|}{Single-Gate} & \multicolumn{4}{c|}{Double-Gate}\\
\cline{2-9}         
\multicolumn{1}{c|}{} & \makecell{$I_{ON}$\\(mA/$\mu$m)} & \makecell{$\rho_{ToB}$\\(cm$^{-2}$)} 
& \makecell{SS$_{15nm}$\\(mV/dec)}  & \makecell{SS$_{5nm}$\\(mV/dec)} &
\makecell{$I_{ON}$\\(mA/$\mu$m)} & \makecell{$\rho_{ToB}$\\(cm$^{-2}$)} 
& \makecell{SS$_{15nm}$\\(cm/s)} & \makecell{SS$_{5nm}$\\(mV/dec)}\\
\hline
  n-As$_4$           &  4.37 &  1.65e13  &   67.7    &  339.7  &  9.06   &  2.66e13  &  61.3  &  229.7  \\
\hline
  p-As$_4$           &  5.12 &  1.47e13  &   73.2    &  689.3  &  10.19  &  2.38e13  &  64.4  &  286.5  \\
\hline
  n-As$_8$Ge$_4$     &  5.12 &  1.54e13  &   71.8    &  254.4  &  11.42  &  2.48e13  &  61.3  &  159.2  \\
\hline
  p-C$_2$N$_4$Pb$_2$ &  4.08 &  1.69e13  &   68.8    &  159.5  &  9.28   &  2.66e13  &  61.6  &  114.2  \\
\hline
  n-GeS              &  4.45 &  1.77e13  &   67.8    &  156.3  &  8.4   &  2.76e13  &  62  &  111  \\
\hline
  p-GeS              &  4.08 &  1.67e13  &   65.8    &  168.4  &  8.67   &  2.79e13  &  61.1  &  119.7  \\
\hline
  n-O$_6$Sb$_4$      &  4.47 &  1.61e13  &   67.6    &  207.7  &  10.73  &  2.61e13  &  62.8  &  136  \\
\hline
  n-P$_4$            &  4.90 &  1.49e13  &   69.3    &  226.9  &  9.44   &  2.36e13  &  62.7  &  161.5  \\
\hline
  p-P$_4$            &  5.21 &  1.58e13  &   70.4    &  231.4  &  10.72  &  2.5e13  &  63.6  &  155.5  \\
\hline
\end{tabular}
}
\caption{ON-state current ($I_{ON}$), charge at the top-of-the-barrier
  ($\rho_{ToB}$), sub-threshold swing at $L_g$=15 nm (SS$_{15nm}$) and
  $L_g$=5 nm (SS$_{5nm}$) for selected 2-D materials with a high $n$-
  and/or $p$-type current, with a single- and a double-gate
  architecture. The $I_{ON}$ and $\rho_{ToB}$ quantities are extracted
  at $L_g$=15 nm.}
\label{tab:2}
\end{table}

\end{document}